\documentclass[10pt,conference]{IEEEtran}
\IEEEoverridecommandlockouts
\usepackage{cite}
\usepackage{amsmath,amssymb,amsfonts}
\usepackage{algorithmic}
\usepackage{graphicx}
\usepackage{textcomp}
\usepackage{xcolor}
\def\BibTeX{{\rm B\kern-.05em{\sc i\kern-.025em b}\kern-.08em
    T\kern-.1667em\lower.7ex\hbox{E}\kern-.125emX}}

\usepackage{graphicx}
\usepackage{algorithmic}
\usepackage{textcomp}
\usepackage{xcolor}
\usepackage{etoolbox}
\usepackage{multirow}
\usepackage{url}
\usepackage{multirow,booktabs}
\usepackage{subcaption}
\usepackage{wrapfig}
\usepackage{float}
\usepackage{float}
\usepackage{tcolorbox}
\usepackage[flushleft]{threeparttable}
\usepackage{relsize}
\usepackage[inline]{enumitem}
\usepackage{amsmath}

\usepackage{algorithm2e}

\restylefloat{figure}

\makeatletter
\newcommand{\linebreakand}{%
  \end{@IEEEauthorhalign}
  \hfill\mbox{}\par
  \mbox{}\hfill\begin{@IEEEauthorhalign}
}
\makeatother

\newcommand{\revise}[1]{{\color{black}{#1}}}

\newcommand{\ourapp}{\textsc{SAFE}}

\newcommand{\quotes}[1]{``#1''}

\newcommand{\rqone}{How does our \ourapp~approach compare to effective and state-of-the-art deep learning-based vulnerability detection baseline methods on real-world, complex, and diverse source code datasets?}

\newcommand{\rqtwo}{How do the enhancing semantic and syntactic relationships (utilizing the distilled knowledge from the teacher models) affect our \ourapp~approach performance?}

\newcommand{\rqthree}{How do different types of hierarchical code structures utilized in the Teacher-B model affect our \ourapp~approach performance?}

\begin{document}

\title{SAFE: Advancing Large Language Models in Leveraging Semantic and Syntactic Relationships for Software Vulnerability Detection}

\author{
    \IEEEauthorblockN{Van Nguyen*\thanks{*Corresponding Author (van.nguyen1@monash.edu). Van Nguyen is a Postdoctoral Research Fellow at the Department of Software Systems and Cybersecurity at Monash University, Australia. Additionally, he is an Affiliate at CSIRO’s Data61, Australia.}}
    \IEEEauthorblockA{\textit{Monash University, Australia}\\
    \textit{CSIRO's Data61, Australia}}
    \and
    \IEEEauthorblockN{Surya Nepal}
    \IEEEauthorblockA{\textit{CSIRO's Data61, Australia}}
    \and
    \IEEEauthorblockN{Tingmin Wu}
    \IEEEauthorblockA{\textit{CSIRO's Data61, Australia}}
    \linebreakand
    \IEEEauthorblockN{Xingliang Yuan}
    \IEEEauthorblockA{\textit{Melbourne University, Australia}}
    \and
    \IEEEauthorblockN{Carsten Rudolph}
    \IEEEauthorblockA{\textit{Monash University, Australia}}
}

\maketitle
\thispagestyle{plain}
\pagestyle{plain}

\begin{abstract}

Software vulnerabilities (SVs) have emerged as a prevalent and critical concern for safety-critical security systems. This has spurred significant advancements in utilizing AI-based methods, including machine learning and deep learning, for software vulnerability detection (SVD). While AI-based methods have shown promising performance in SVD, their effectiveness on real-world, complex, and diverse source code datasets remains limited in practice. 

To tackle this challenge, in this paper, we propose a novel framework that enhances the capability of large language models to learn and utilize semantic and syntactic relationships from source code data for SVD. As a result, our approach can enable the acquisition of fundamental knowledge from source code data while adeptly utilizing crucial relationships, i.e., semantic and syntactic associations, to effectively address the software vulnerability detection (SVD) problem.

The rigorous and extensive experimental results on three real-world challenging datasets (i.e., ReVeal, D2A, and Devign) demonstrate the superiority of our approach over the effective and state-of-the-art baselines. In summary, on average, our \ourapp~approach achieves higher performances from 4.79\% to 9.15\% for F1-measure and from 16.93\% to 21.70\% for Recall compared to the baselines across all datasets used.

\end{abstract}

\section{Introduction}\label{sec:introduction}

Software vulnerabilities (SVs) are specific flaws or oversights in software programs that allow attackers to exploit the code base, potentially enabling them to carry out dangerous activities such as exposing or altering sensitive information, disrupting, degrading, or destroying systems, or taking control of programs or computer systems \cite{Dowd2006, fu2024ai}. These vulnerabilities are widespread and pose significant security risks due to the pervasive use of computer software. Over the years, the severity of threats posed by SVs has escalated, causing considerable damage to companies and individuals. This worsening situation has necessitated the creation of automated advanced methods and tools that can efficiently and effectively detect SVs with minimal human intervention \cite{vannguyen2019dan, van-dual-dan-2020, velvet2022, fu2022vulrepair, fu2024vision}.

Software vulnerability detection (SVD) plays a vital role in software engineering to ensure the security and integrity of software applications \cite{Dowd2006, Lin2020, Hanif2021, nguyenicvh2021, fu2022linevul, liu2023refining, fu2023chatgpt, nguyen2024deep}. Identifying vulnerable programs or functions is an essential aspect of the security engineering process, allowing security professionals to efficiently allocate resources and address critical vulnerabilities during development and testing phases. This process ultimately enhances the security and reliability of software applications. To meet this demand, a variety of vulnerability detection systems and methods have been proposed and implemented, ranging from open-source to commercial tools and from manual to automated methods \cite{Neuhaus:2007:PVS,shin2011evaluating,Grieco2016,VulDeePecker2018, Duan2019, Cheng2019, wattanakriengkrai2020predicting, vandam2p2024}.

Most previous research in software vulnerability detection (SVD), such as \cite{yamaguchi2011vulnerability, shin2011evaluating, Li2016:VAV, Grieco2016, KimWLO17}, has relied primarily on handcrafted features selected manually by domain experts, whose experience may be outdated and biased. These handcrafted features often lack generalizability. For instance, features that perform well in one software project may not be effective in others \cite{Zimmermann2009}. To reduce the reliance on handcrafted features, the use of automatic features in SVD through deep learning-based techniques has been explored \cite{VulDeePecker2018, Dam2018, Li2018SySeVR, nguyen2022info, fu2023lqsvd, fu2024aibughunter, vandam2p2024}. The deep learning-based methods demonstrate the advantages of automatic features over handcrafted ones in addressing the SVD problem.

Recently, there has been a growing exploration of large language models for software vulnerability detection \cite{CodeBERT2020, Guographcodebert2021, wang-etal-2021-codet5, gao2023far,Michael2023VulExplainer,fu2023chatgpt,yao2024}. These studies have investigated and shown the promising capability of such models to extract fundamental knowledge from source code data, facilitating effective problem-solving. However, large language models often struggle with complex datasets due to two primary reasons: Firstly, the lack additional contextual information beyond the source code data itself, as these models often struggle to perform well without ample supplementary information elucidating the intricacies of the source code. Secondly, they face limitations in effectively learning and leveraging the semantic and syntactic relationships embedded within the source code data when applied to downstream SVD tasks.

It is noteworthy that the capability to simultaneously learn fundamental knowledge and leverage both semantic and syntactic relationships from source code data remains limited in most current machine learning and deep learning approaches. \revise{This limitation significantly impacts their performance in software vulnerability detection, particularly when confronted with real-world, complex, and diverse source code datasets.} These datasets often reflect realistic vulnerability scenarios, lack of additional context beyond the source code data itself, and exhibit diversity between the testing and training datasets, further exacerbating the challenges faced by current machine learning and deep learning approaches. As a consequence, the performance of existing SVD methods on these types of datasets remains limited.

\revise{To effectively mitigate this problem, in this paper,} we introduce an innovative deep learning-based approach for software vulnerability detection. Our approach enhances the capability of large language models to proficiently learn and leverage semantic and syntactic relationships from source code data by incorporating a distillation mechanism. This fusion results in an innovative method that can not only acquire the  fundamental knowledge from source code data but also adeptly utilize crucial relationships, i.e., the semantic and syntactic associations, to address the SVD problem.

Particularly, our proposed \ourapp~approach consists of two phases. \textbf{In the first phrase}, we train and utilize two lightweight and effective deep-leaning-based models (denoted as teacher-A and teacher-B) for delicately learning the semantic and syntactic relationships from the source code data, respectively, when addressing the software vulnerability detection problem. \textbf{In the second phase}, we distil the knowledge from the trained-teacher-A and trained-teacher-B models (obtained from the first phase) to a Student-S model (i.e., built based on a large language model backbone utilizing either of two effective and popular pre-trained large language models RoBERTa \cite{LiuRoBERTa2019} or CodeT5 \cite{wang-etal-2021-codet5}) using the distillation mechanism to enhance its capability in effectively learning and leveraging the semantic and syntactic relationships from the source code data. This elegant integration allows the Student-S model to not only acquire the fundamental knowledge from source code data (i.e., as a consequence of being built from a pre-trained large language model) but also adeptly utilize the data crucial semantic and syntactic association, for effectively tackling the software vulnerability detection problem.

Through rigorous and extensive experimental evaluation of our \ourapp~approach on three real-world challenging source code datasets, ReVeal \cite{chakraborty2020deep}, D2A \cite{d2a2021}, and Devign \cite{ReGVD2022}, we address the following two main research questions:

\begin{itemize}
  \item \textbf{(RQ1) How does our proposed \ourapp~approach compare to effective and state-of-the-art deep learning-based vulnerability detection baselines on real-world challenging source code datasets? \\
  Results.} The experimental results on three real-world, complex, and diverse datasets demonstrate the significant advancement of our \ourapp~approach compared to the baselines, particularly in the F1-measure (the harmonic mean of Recall and Precision) and Recall metrics. In particular, on average, our \ourapp~approach achieves significantly higher performances from \textit{4.79\% to 9.15\% for the F1-measure and from 16.93\% to 21.70\% for Recall} compared to the baselines across all datasets used. In software vulnerability detection (SVD), F1-measure can be considered the most important metric, with Recall having a higher priority than Precision \cite{ami2023false}.
  \vspace{1mm}
  \item \textbf{(RQ2) How do (i) enhancing semantic and syntactic relationships via the distillation mechanism and (ii) the types of hierarchical code structures (i.e., AST and DFG) used in the teacher-B model affect the performance of our \ourapp~approach?\\
  Results.}
  The experimental results from these ablation studies demonstrate the effectiveness and advancement of our \ourapp~approach in enhancing the capability of large language models to learn and leverage semantic and syntactic relationships from source code data to address the SVD problem. This combination of capabilities enables our \ourapp~approach not only to acquire fundamental knowledge from source code data but also to adeptly utilize crucial relationships within the source code, effectively solving the SVD problem.
\end{itemize}

\vspace{1mm}
\textbf{Novelty and Contributions} The contributions of our paper are as follows:

\begin{itemize}

  \item We proposed a novel deep learning-based approach, namely \ourapp, for solving the software vulnerability detection (SVD) problem. Our \ourapp~approach enhances the capability of large language models to learn the semantic and syntactic relationships from source code, for effectively tackling the SVD problem. To the best of our knowledge, our \ourapp~approach is among the first that can be able to leverage large language models for fundamental knowledge extraction while effectively utilizing both semantic and syntactic relationships from source code to tackle the SVD problem.
  \vspace{1mm}
  \item We conducted rigorous and extensive experiments on three real-world challenging source code datasets (i.e., ReVeal \cite{chakraborty2020deep}, D2A \cite{d2a2021}, and Devign \cite{ReGVD2022}), \textit{where the performance of existing SVD methods remains limited}. The experimental results show that our \ourapp~approach significantly outperforms seven effective and state-of-the-art baselines (i.e., TextCNN \cite{textCNNKim2014}, RoBERTa \cite{LiuRoBERTa2019}, Devign \cite{DevignZhou2019}, CodeBERT \cite{CodeBERT2020}, GraphCodeBERT \cite{Guographcodebert2021}, CodeT5 \cite{wang-etal-2021-codet5}, and ReGVD \cite{ReGVD2022}), particularly in F1-measure and Recall, the key and prioritized metrics in software vulnerability detection \cite{VulDeePecker2018,chakraborty2020deep,DevignZhou2019,d2a2021,ami2023false}.
\end{itemize}
\section{Related Work}\label{sec:related_work}

AI-based approaches have been widely proposed for Software vulnerability detection (SVD), ranging from utilizing handcrafted features manually selected by domain experts \cite{yamaguchi2011vulnerability, shin2011evaluating, Li2016:VAV, Grieco2016, KimWLO17}, to leveraging automatic feature learning through deep learning-based methods \cite{VulDeePecker2018,jun_2018, Dam2018, Li2018SySeVR, Duan2019, Cheng2019, Zhuang2020, ReGVD2022, nguyen2022info, nguyen2022cross}, showcasing notable advancements in the field. For some examples, \cite{Dam2018} employed a deep neural network to transform sequences of code tokens to vectorial features that are further fed to a separate classifier; whereas \cite{VulDeePecker2018} combined the learning of the vector representation and the training of the classifier in a deep network. Advanced deep net architectures have further been investigated for addressing the SVD problem. The study in \cite{Rebecca2018} combined both recurrent neural networks (RNNs) and convolutional neural networks (CNNs) for feature extraction from the embedded source code representations while \cite{Zhuang2020,ReGVD2022} proposed deep learning-based models, namely TMP and ReGVD, respectively, for SVD based on graph neural networks \cite{KipfW16,Gated2016,GCNN2017}.

The works from \cite{CodeBERT2020,Guographcodebert2021,wang-etal-2021-codet5,gao2023far,fu2023chatgpt,yao2024} have studied and utilized large language models (LLMs) for software vulnerability detection. CodeBERT, GraphCodeBERT, and CodeT5 \cite{CodeBERT2020,Guographcodebert2021,wang-etal-2021-codet5} are pre-trained large language models for multiple programming languages and relevant tasks such as code search, code completion, code summarization. They are also fine-tuned and applied for other code-related downstream tasks, e.g., software vulnerability detection and obtained promising performances in solving the problem. \cite{gao2023far,fu2023chatgpt,yao2024} investigated the performance of ChatGPT and LLaMA in addressing the software vulnerability detection problem. These studies revealed the limitations of these well-known large language models due to the absence of explanatory context in downstream datasets and the complexity of downstream tasks, which impedes their adoption. \revise{Furthermore, they suggested that beyond the source code data itself, obtaining more relevant context to elucidate the intricacies of the source code could significantly aid the models in making better predictions}.
\section{The proposed \ourapp~approach}\label{sec:framework}

\subsection{\textbf{The problem statement}}\label{sec:problemstatement}
We denote $D$ as a real-world source code dataset consisting of $\{(X_{1}, Y_{1}),\dots,(X_{N}, Y_{N})\}$ where $X_i$ is a source code data sample (i.e., a function or a program) while $X_i$'s vulnerability label $Y_{i}\in\{0,1\}$ (i.e., $0$: non-vulnerable and $1$: vulnerable). In this paper, we study the problem of software vulnerability detection, aiming to automatically predict the vulnerability label $Y_{i}$ for each source code sample $X_{i}$.

\vspace{0mm}
\subsection{\textbf{Methodology}}

In what follows, we present the details of how our proposed \ourapp~approach works to enhance the capability of large language models in learning and leveraging the semantic and syntactic relationships from source code data (utilizing the distillation mechanism) for effectively addressing the software vulnerability detection (SVD) problem. 

A detailed and comprehensive visualization of our \ourapp~approach is depicted in Figure \ref{fig:framework}. In particular, our \ourapp~approach's framework employs a large language model backbone as the student model (namely Student-S), which is trained to emulate the behaviors of two specialized teacher models via the distillation mechanism. These teachers focus on extracting and learning semantic and syntactic relationships from source code data when solving the software vulnerability detection problem. Moreover, the student model, based on a large language model backbone, can also autonomously learns directly from the source code data itself in utilizing its capability in learning the data fundamental knowledge. This facilitates data representation learning, effectively aiding in solving the problem. In short, our \ourapp~approach comprises two phases as follows:

\begin{figure*}[ht]
\vspace{-3mm}
\begin{centering}
\includegraphics[width=0.91\textwidth]{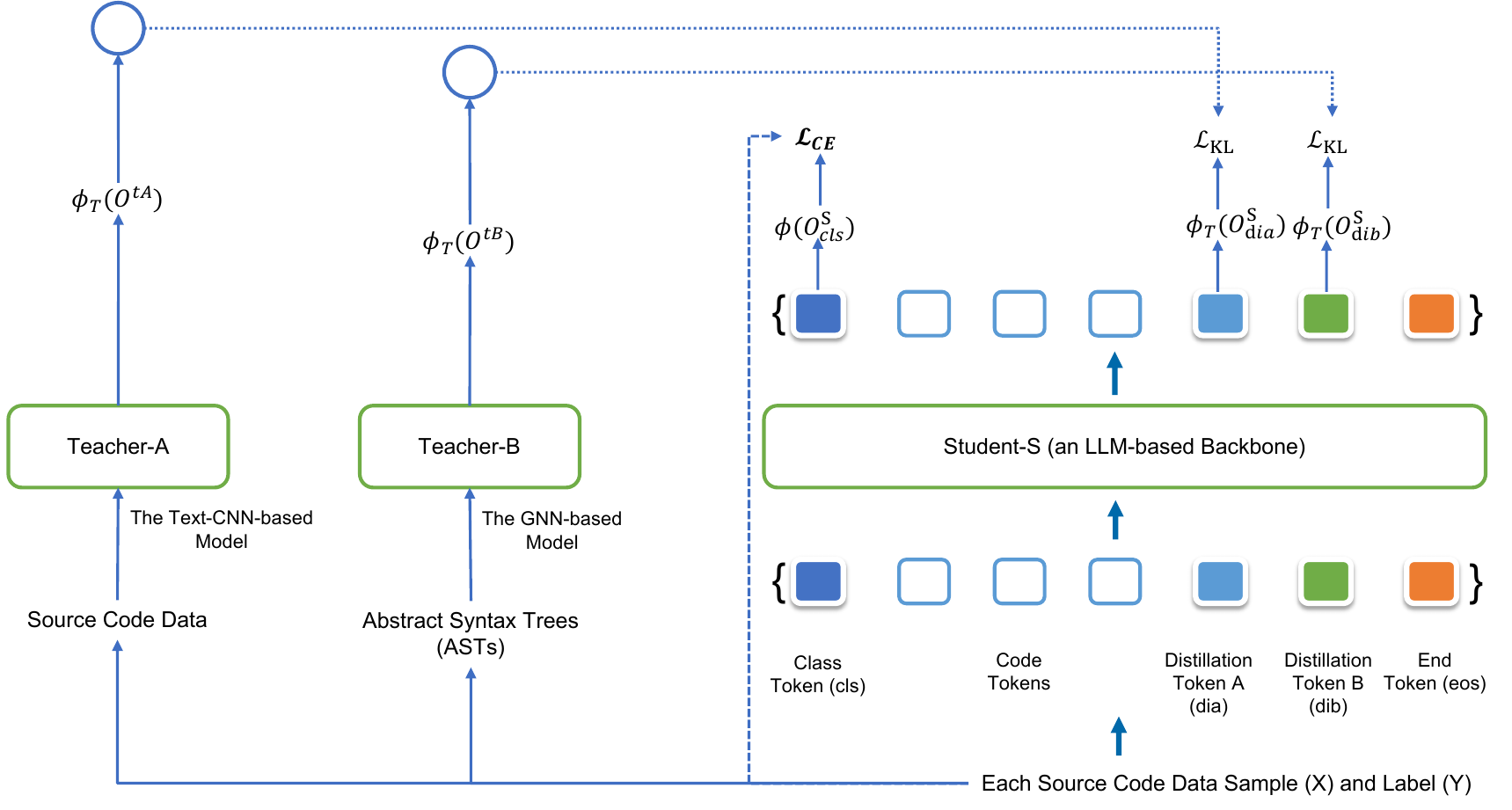}
\par\end{centering}\vspace{0.5mm}
\caption{A visualization of our \ourapp~approach in enhancing the capability of large language models to learn the semantic and syntactic relationships in the source code data. This fusion enables our \ourapp~approach to not only acquire fundamental knowledge from the source code (as a consequence of being built from a pre-trained large language model) but also adeptly utilize crucial semantic and syntactic associations, effectively improving its ability to address the SVD problem.}\label{fig:framework}
\vspace{-3mm}
\end{figure*}%
\vspace{-0mm}

\vspace{0mm}
\subsubsection{\textbf{Training two teacher models focusing on learning the semantic and syntactic relationships of source code data}}

In the first phase, we train two deep learning-based specialized teacher models (i.e., namely teacher-A and teacher-B) focusing on extracting semantic and syntactic relationships, respectively, from source code when addressing the software vulnerability detection (SVD) problem. 

\vspace{0mm}
\paragraph{\textbf{Teacher-A}}
Teacher-A is dedicated to learning semantic relationships to tackle the software vulnerability detection problem. In this case, we treat each source code sample as a sequence of code tokens and then employ the TextCNN model for this solving the problem. We choose this model mainly because it is widely recognized as a lightweight and effective approach for processing sequential data and learning the data representations \cite{textCNNKim2014}. Given a source code dataset $D$ = $\{(X_{1}, Y_{1}),\dots,(X_{N}, Y_{N})\}$ as  mentioned in section \ref{sec:problemstatement}, the teacher-A model is trained to minimize the objective function Eq. (\ref{eq:teacher-A}) as follows: 

\vspace{-3mm}
{
\begin{equation}
\underset{\alpha}{\text{min}} \ \frac{1}{N}\sum_{i=1}^{N} \ \mathcal{L}(f(X_{i},\alpha),Y_{i})\label{eq:teacher-A}
\end{equation}
}
\vspace{-3mm}

where $\mathcal{L}$ represents the cross-entropy loss function (i.e., measuring the discrepancy between $f(X_{i},\alpha)$ and $Y_{i}$), and $f$ denotes the teacher-A model with the parameter $\alpha$.

\vspace{0mm}
\paragraph{\textbf{Teacher-B}}
On the other hand, teacher-B will concentrate on learning syntactic relationships (or code structures) from the source code data when tackling the software vulnerability detection problem. In this analysis, we delve into the hierarchical structure of the source code samples, focusing on the abstract syntax tree (AST) as it furnishes comprehensive syntax details of the source code data. To capture the AST of each source code sample, we leverage Treesitter \cite{Treesitter}, an effective and commonly used parser generator tool. Subsequently, we convert the AST into a sequence, preserving all structural information intact \cite{guo2022unixcoder}, which facilitates more efficient processing of the source code data. We then employ a graph neural network, widely recognized for its ability to learn syntactic information from data, as used in \cite{ReGVD2022}, to capture the syntactic relationships within the source code data when solving the SVD problem.

Given a source code dataset $D$ = $\{(X_{1}, Y_{1}),\dots,(X_{N}, Y_{N})\}$ as  mentioned in section \ref{sec:problemstatement}, we use Treesitter \cite{Treesitter} to obtain the corresponding AST structure of each source code data sample $X_{i}$ before using \cite{guo2022unixcoder} to convert this AST structure into a corresponding sequence denoted as $Xast_{i}$, and then teacher-B model is trained to minimize the objective function Eq. (\ref{eq:teacher-B}) as follows: 

\vspace{-3mm}
{
\begin{equation}
\underset{\beta}{\text{min}} \ \frac{1}{N}\sum_{i=1}^{N} \ \mathcal{L}(g(Xast_{i},\beta),Y_{i})\label{eq:teacher-B}
\end{equation}
}
\vspace{-3mm}

where $\mathcal{L}$ represents the cross-entropy loss function (i.e., measuring the discrepancy between $g(Xast_{i},\beta)$ and $Y_{i}$) while $g$ denotes the teacher-B model with the parameter $\beta$.

\vspace{0mm}
\subsubsection{\textbf{Training a student model utilizing a large language model backbone with the distillation mechanism to enhance its ability to learn semantic and syntactic relationships in source code data}}

In the second phase, we train a student model (denoted as Student-S) utilizing a large language model backbone to address the software vulnerability detection problem. Specifically, we sequentially leverage two effective and popular pre-trained models, CodeT5 \cite{wang-etal-2021-codet5} and RoBERTa \cite{LiuRoBERTa2019}, as the backbone for the Student-S model.

Given a source code data sample $X$, we tokenize $X$ into a sequence of code tokens $[{x}_{1},...,{x}_{l}]$ using the byte pair encoding (BPE) algorithm \cite{sennrich2016neural} (i.e., which is commonly used in large language models, e.g., GraphCodeBERT \cite{Guographcodebert2021}). We then truncate and do padding to let $l=512$ tokens. Next, we embed each code token ${x}_{i}$ in $X$ and each structure token ${xast}_{i}$ in $Xast$ into corresponding representation vectors $\boldsymbol{w}_{i}$ and $\boldsymbol{a}_{i}$, respectively, using the embedding layer from the large language model backbone architecture (i.e., CodeT5 or RoBERTa) used in the Student-S model. Afterward, the sequence of code token representation vectors $W$ = $[\boldsymbol{w}_{1},...,\boldsymbol{w}_{l}]$ will be used as input for both the trained-teacher-A and student-S models. The trained-teacher-B model takes the sequence of structure token representation vectors $A$ = $[\boldsymbol{a}_{1},...,\boldsymbol{a}_{l}]$ as input.

It is important to note that in our framework, four special tokens, i.e., $[cls]$, $[dia]$, $[dib]$, and $[sep]$, are added during the tokenization. The class embedding ($[cls]$) is used to learn the representation of the input source code data sample $X$, which will be utilized by a classification head to assess the vulnerability $Y$ of $X$. Additionally, the $[dia]$ and $[dib]$ tokens are inserted before the $[sep]$ token to distill knowledge from the trained-teacher-A and trained-teacher-B models, respectively. These distillation embeddings ($[dia]$ and $[dib]$) enable the Student-S model to learn from the outputs of the trained-teacher-A and trained-teacher-B models, indirectly enhancing its capability to learn and leverage semantic and syntactic relationships from source code data. This process is similar to regular distillation, while remaining complementary to the class embedding $[cls]$. This elegant integration allows the Student-S model to not only acquire fundamental knowledge (i.e., as a consequence of being built from a pre-trained large language model) from source code data $X$ but also can adeptly utilize the data crucial semantic and syntactic associations, facilitating its ability to predict the corresponding vulnerability $Y$ of $X$.

Denoting the output of the embedding layer from the backbone large language model used in the Student-S model as $H^{0}$=$[\boldsymbol{h}_{1},...,\boldsymbol{h}_{l}]$. The embedding vectors $H^{0}$ then will be gone
through $q$ multi-head self-attention encoder layers of the large language model backbone's encoder used in the Student-S model (i.e., in our study, we utilize either of two effective and popular pre-trained large language models, RoBERTa or CodeT5, using their base versions with $q=12$) to learn the representation of each token from the input $X$: $H^{z}$ = $E^{z}(H^{z-1})$ with $z \in {1,...,q}$.

In particular, each encoder layer $E^{z}$ consists of a multi-head self-attention operation (i.e., it allows the model to jointly attend to information from different representation sub-spaces at different positions of the input) followed by a feed-forward neural network. Firstly, encoder layer $E^{z}$ takes the $H^{z-1}$ as input to the multi-head self-attention operation (MHA) followed by a layer-norm operation (LN):

\vspace{-3mm}
{
\begin{equation}
M^{z} = LN(MHA(H^{z-1}) + (H^{z-1}))\label{eq:Ai}
\end{equation}
}
\vspace{-3mm}

The output $M^{z}$ will then be gone through the  fully connected feed-forward network (FFN) with a layer-norm operation (LN) to result in $H^{z}$:

\vspace{-3mm}
{
\begin{equation}
H^{z} = LN(FFN(M^{z}) + (M^{z}))\label{eq:Hi}
\end{equation}
}
\vspace{-3mm}

At the last multi-head self-attention encoder layer $E^{q}$, we possess the token representation embeddings $H^{q}$ consisting of the class token embedding ${H^{q}}_{cls}$ and the distillation token embeddings ${H^{q}}_{dia}$ and ${H^{q}}_{dib}$. We then feed these representation embeddings to three fully connected feed-forward neural networks to work out the class token logits ${O^S}_{cls}$ and the distilled token logits ${O^S}_{dia}$  and ${O^S}_{dib}$.

\vspace{0mm}
\paragraph{\textbf{Knowledge distillation}}

Neural networks typically produce class probabilities using the softmax function on the output layer of the networks. This function converts the output logit, $o_{k}$, computed for each class into a probability, $p_{k}$, by comparing $o_{k}$ with the other logits:

\vspace{-2mm}
{
\begin{equation}
p_{k} = \frac{exp(o_{k}/T)}{\sum_{j}exp(o_{j}/T)}\label{eq:distillation}
\end{equation}
}
\vspace{-2mm}

where, $T$ is a temperature parameter typically set to $1$. It is know that using a higher value for $T$ produces a softer probability distribution over classes. 

Knowledge distillation \cite{hinton2015distilling,tejankar2021isd,tuannguyen2022,li2022curriculumt} is a technique used to transfer knowledge from one or more teacher models to a student model. In its simplest form, the student model is trained on a transfer set of data using soft target probability distributions produced by the teacher models, employing a temperature $T$ in the softmax function.
The student model is trained with the same temperature $T$, but uses a temperature of 1 after training. Inspired by \cite{hinton2015distilling}, we transfer knowledge from the trained-teacher-A and trained-teacher-B models to the Student-S model by minimizing the Kullback-Leibler (KL) divergence between the softmax outputs of the teacher models and the student model's distilled tokens. Particularly, let $O^{tA}$ and $O^{tB}$ be the logit output values of the trained-teacher-A and trained-teacher-B models for a given source code data sample $X$ with its corresponding vulnerability label $Y$. The Student-S model is trained to minimize the following objective function:

\vspace{-3mm}
{
\begin{equation}
\begin{split}
\mathcal{L}_{S} & = \gamma\mathcal{L}_{CE}(\phi({O^S}_{cls}),Y)+ \delta\mathcal{L}_{KL}(\phi_{T}({O^S}_{dia}),\phi_{T}(O^{tA})) \\
& + \eta\mathcal{L}_{KL}(\phi_{T}({O^S}_{dib}),\phi_{T}(O^{tB}))
\label{eq:student-s}
\end{split}
\end{equation}
}
\vspace{-3mm}

Where $\phi_{T}$ is the softmax function with the temperature $T$. When the temperature $T$ is set to 1, we denote the softmax function as $\phi$. The trade-off hyper-parameters $\gamma$, $\delta$, and $\eta$ (where $\gamma + \delta + \eta = 1$) used to set the weights of utilizing the pre-trained large language model backbone in the Student-S model as well as enhancing the semantic and syntactic relationships transferred from the trained-teacher-A and trained-teacher-B models.

\vspace{0mm}
\subsubsection{\textbf{Inference (testing) phase}}
Given a source code data sample $X$ from the testing set $D_{test}$, we based on the representation of the class token $[cls]$ (i.e., ${O^S}_{cls}$) of the trained Student-S model in our proposed \ourapp~approach to predict the vulnerability $Y$ of data $X$.

\vspace{-3mm}
{
\begin{equation}
\begin{split}
\Tilde{\boldsymbol{p}} & = \phi({O^S}_{cls}) \\
\Tilde{Y} & = argmax(\Tilde{\boldsymbol{p}}) = argmax_{j}(\Tilde{p_{j}}) \\
\label{eq:testing}
\end{split}
\end{equation}
}
\vspace{-5mm}

The comprehensive algorithm of our \ourapp~approach is exhibited in Algorithm \ref{alg:The-training-algorithm-soda}.

\RestyleAlgo{ruled}
\begin{algorithm*}[htbp]
\vspace{1mm}
\DontPrintSemicolon
\LinesNumbered

\KwIn{
A real-world source code dataset $D=\left\{ \left(X_{1},Y_{1}\right),\dots,(X_N,Y_N)\right\}$ where each source code data sample $X_{i}$ consisting of $l$ code tokens from $\mathbf{x}_{1}$ to $\mathbf{x}_{l}$ while its vulnerability $Y_{i}\in\left\{ 0,1\right\} $ (i.e., $1$: vulnerable and $0$: non-vulnerable).

The dataset $D$ is split into three sets, as investigated and used in the baselines (i.e., in our study, we simply use the split version widely recognized by the community): $D_{train}$ for training the model, $D_{val}$ for saving the best model during the training process, and $D_{test}$ for testing the model's performance after the training phase. We denote the number of training epochs as $eps$ and the mini-batch size as $m$.
}
\BlankLine
\textbf{Phase (1)}
\BlankLine
We initialize the parameters $\alpha$ and $\beta$ of the teacher-A (i.e., $f(.,\alpha)$) and teacher-B (i.e., $g(.,\beta)$) models, respectively.
\BlankLine
\For{$t=1$ to $eps$}
{
Choose a mini-batch of source code samples denoted by $\{(X_{i},Y_{i})\}_{i=1}^{m}$ from $D_{train}$ used for the teacher-A model.
\BlankLine

Use Treesitter \cite{Treesitter} to obtain the corresponding ASTs, then convert these data into sequences while preserving all structural information intact \cite{guo2022unixcoder}. This approach facilitates more efficient processing of the code data structures, denoted by ${(Xast_{i},Y_{i})}_{i=1}^{m}$, used for the teacher-B model.

\BlankLine
Update the parameter $\alpha$ of the teacher-A model via minimizing the cross-entropy loss function $\underset{\alpha}{\text{min}} \ \frac{1}{m}\sum_{i=1}^{m} \ \mathcal{L}(f(X_{i},\alpha),Y_{i})$ as mentioned in Eq. (\ref{eq:teacher-A}) using the Adam optimizer \cite{KingmaB14}.
\BlankLine
Update the parameter $\beta$ of the teacher-B model via minimizing the cross-entropy loss function $\underset{\beta}{\text{min}} \ \frac{1}{m}\sum_{i=1}^{m} \ \mathcal{L}(g(Xast_{i},\beta),Y_{i})$ as mentioned in Eq. (\ref{eq:teacher-B}) using the Adam optimizer \cite{KingmaB14}.
\BlankLine
}

\textbf{Note that}, after phase (1), we have two effective teacher models dedicated to learning the semantic and syntactic relationships from the source code data, respectively, when tackling the software vulnerability detection (SVD) problem.

\BlankLine
\textbf{Phase (2)}
\BlankLine
We set the large language model backbone for the Student-S model (i.e., CodeT5 or RoBERTa, two effective and popular pre-trained large language models). We denote the parameter of the Student-S model as $\theta$.
\BlankLine
\For{$t=1$ to $eps$}
{
Choose a mini-batch of source code samples denoted by $\{(X_{i},Y_{i})\}_{i=1}^{m}$ from $D_{train}$ used for the trained-teacher-A and Student-S models.
\BlankLine
Use Treesitter \cite{Treesitter} to obtain the corresponding ASTs, then convert these data into sequences while preserving all structural information intact \cite{guo2022unixcoder}. This approach facilitates more efficient processing of the code data structures, denoted by ${(Xast_{i},Y_{i})}_{i=1}^{m}$, used for the trained-teacher-B model.
\BlankLine
Update the parameter $\theta$ of the Student-S model via minimizing the objective function $\mathcal{L}_{S} = \gamma\mathcal{L}_{CE}(\phi({O^S}_{cls}),Y)+ \delta\mathcal{L}_{KL}(\phi_{T}({O^S}_{dia}),\phi_{T}(O^{ta})) + \eta\mathcal{L}_{KL}(\phi_{T}({O^S}_{dib}),\phi_{T}(O^{tb}))$ as mentioned in Eq. (\ref{eq:student-s}) using the Adam optimizer \cite{KingmaB14}. 

\textbf{Note that}: Via this objective function, we distil the knowledge from the trained-teacher-A and trained-teacher-B for enhancing the capability of the Student-S model to learn semantic and syntactic relationships in source code data. This fusion allows the Student-S model to not only acquire the fundamental knowledge from source code data (i.e., as a consequence of being built from a pre-trained large language model) but also adeptly utilize the data crucial semantic and syntactic association, for effectively tackling the software vulnerability detection problem.
}

\BlankLine
\textbf{Testing phase}
\BlankLine
Given a source code data $X_{i}$ from the testing set $D_{test}$, we based on the representation of the class token $[cls]$ in the trained Student-S model to predict the data $X_{i}$' vulnerability $Y_{i}$ as mentioned in Eq. (\ref{eq:testing}).

\BlankLine
\KwOut{The trained Student-S model capable for predicting the vulnerability (vulnerable or non-vulnerable) of source code data samples.}\caption{The algorithm of our \ourapp~approach in enhancing the capability of large language models in learning the semantic and syntactic relationships inside source code data, ultimately
improving addressing the software vulnerability detection (SVD) problem.\label{alg:The-training-algorithm-soda}}
\vspace{0mm}
\end{algorithm*}

\vspace{0mm}

\section{Experiments}\label{sec:experiments}

\subsection{\textbf{Experimental Designs}}

The key goal of our experiments is to evaluate our \ourapp~method and compare it with effective and state-of-the-art baselines for solving the SVD problem on real-world, complex, and diverse source code datasets. Below, we present the research questions of our experiments.

\vspace{1mm}
\textbf{(RQ1) \rqone}

The serious problem of software vulnerabilities has spurred the development of AI-based methods, including machine learning and deep learning approaches, to address this issue. While these AI-based methods have shown promising results in software vulnerability detection (SVD), their performance remains limited when applied to real-world, complex, and diverse source code datasets. These datasets often reflect realistic vulnerability scenarios, lack additional context beyond the source code itself, and exhibit diversity between the testing and training datasets. The main reason for this limitation is likely that these models struggle to simultaneously learn fundamental knowledge and leverage both semantic and syntactic relationships from source code data, compromising their generalization and effectiveness. Recently, the application of large language models for software vulnerability detection has been studied and explored. While these models also show promise, they face challenges with complex source code datasets due to the lack of contextual information beyond the code itself and their limitations in effectively learning and leveraging the semantic and syntactic relationships embedded within the source code from the downstream tasks. 

To this end, in this paper, we propose an innovative deep learning-based approach that enhances the ability of large language models to learn and leverage semantic and syntactic relationships from source code data using the distillation mechanism. This fusion creates an innovative method that not only acquires fundamental knowledge from source code but also adeptly utilizes crucial semantic and syntactic associations, ultimately improving addressing the problem.

Via this research question, we investigate the performance of our \ourapp~approach compared to effective and state-of-the-art baselines when tackling the real-world, complex, and diverse source code datasets (i.e.,  ReVeal \cite{chakraborty2020deep}, D2A \cite{d2a2021}, and Devign \cite{ReGVD2022}). These datasets represent realistic vulnerability scenarios, lack additional context beyond the source code itself, and show significant diversity between testing and training sets.

\vspace{3mm}
\textbf{(RQ2) \rqtwo}

Our \ourapp~approach enhances the capability of large language models (used as the Student-S model in our framework) in learning and leveraging semantic and syntactic relationships from source code data (via the distilled knowledge from two trained-teacher-A and trained-teacher-B models) for effectively solving the software vulnerability detection problem. In this research question, we investigate how the use of enhancing the semantic and syntactic relationships affects the performance of our \ourapp~approach.

\vspace{3mm}
\textbf{(RQ3) \rqthree}

In this research, we investigate the performance of our \ourapp~approach in cases of utilizing different hierarchical structures from the source code data. These include the use of the abstract syntax tree (AST), which furnishes comprehensive syntax details of the source code, and the data flow graph (DFG), which represents the \quotes{where-the-value-comes-from} relationships between variables, in learning the source code data representation for Teacher-B.

\vspace{1mm}
\subsubsection{\textbf{Studied datasets}}\label{sec:studied_data_sets}

We primarily conducted experiments on three real-world challenging source code datasets including ReVeal \cite{chakraborty2020deep}, D2A \cite{d2a2021}, and Devign \cite{ReGVD2022}. The ReVeal dataset comprises real-world vulnerabilities reported by developers and users from the Chromium and Debian projects. For the D2A dataset, we use the D2A Leaderboard Data, which consists of security errors from real-world Libav, OpenSSL, Nginx, Httpd, and Libtiff projects. The Devign dataset is collected from four large C-language open-source projects that are popular among developers and diverse in functionality, i.e., Linux Kernel, QEMU, Wireshark, and FFmpeg.

\vspace{1mm}
For these datasets, \textit{we used the community-recognized split versions of the training, validation, and testing sets, as utilized by baseline methods}. These datasets are well-known as challenging because of reflecting realistic vulnerability scenarios, lacking of additional context beyond the source code data itself, and exhibiting diversity between the testing and training datasets which exacerbate the challenges faced by current software vulnerability detection (SVD) approaches.

\vspace{2mm}
\subsubsection{\textbf{Baseline methods}}

The baselines of our proposed \ourapp~approach are seven effective and state-of-the-art methods for software vulnerability detection including TextCNN \cite{textCNNKim2014}, RoBERTa \cite{LiuRoBERTa2019}, Devign \cite{DevignZhou2019}, CodeBERT \cite{CodeBERT2020}, GraphCodeBERT \cite{Guographcodebert2021}, CodeT5 \cite{wang-etal-2021-codet5}, and ReGVD \cite{ReGVD2022}.
These baselines cover a wide range of techniques and methods for learning crucial semantic and syntactic relationships in source code data to address the SVD problem. They include convolutional neural networks (TextCNN), graph neural networks (Devign and ReGVD), and large language-based models (RoBERTa, CodeBERT, GraphCodeBERT, and CodeT5).

\vspace{1mm}
\subsubsection{\textbf{Measures}}

To measure the performance of our \ourapp~approach and baselines, we use three main metrics used in software vulnerability detection including Recall, Precision, and F1-measure (e.g., \cite{VulDeePecker2018,Li2018SySeVR,vannguyen2019dan,DevignZhou2019,d2a2021}). In the field of software vulnerability detection (SVD), the F1-measure (the harmonic mean of Recall and Precision) can be considered the most important metric, with Recall prioritized over Precision \cite{ami2023false}. Higher values in these metrics indicate better performances.

\vspace{1mm}
\subsection{\textbf{Experimental results}}\label{sec:exp_results}

\textbf{RQ1: \rqone}
\vspace{0mm}
\paragraph{\textbf{Approach}}
We compare the performance of our \ourapp~method with seven effective and state-of-the-art software vulnerability detection (SVD) baselines, i.e., TextCNN \cite{textCNNKim2014}, RoBERTa \cite{LiuRoBERTa2019}, Devign \cite{DevignZhou2019}, CodeBERT \cite{CodeBERT2020}, GraphCodeBERT \cite{Guographcodebert2021}, CodeT5 \cite{wang-etal-2021-codet5}, and ReGVD \cite{ReGVD2022}, on three main metrics including Recall, Precision, and F1-measure.

\paragraph{\textbf{Results}}
The experimental results in Table \ref{tab:mainresults} demonstrate the effectiveness and superiority of our \ourapp~approach over the baselines. In particular, our \ourapp~approach obtains noticeably higher performances in the key and prioritized metrics, i.e., F1-measure and Recall, with a high margin for all datasets (i.e., ReVeal, D2A, and Devign). 

\begin{table}[h]
\vspace{0mm}
\caption{The experimental results for Recall, Precision, and F1-measure of our approach and baselines on the testing set of Reveal, D2A, and Devign datasets. We denote our approach as \ourapp-R and \ourapp-T when using RoBERTa and CodeT5, respectively, as the backbone for the Student-S model. The numbers highlighted in blue represent the improvements of \ourapp-R and \ourapp-T over the second-best baseline.}\label{tab:mainresults}
\vspace{-1mm}
\centering{}
\resizebox{0.99\columnwidth}{!}{
\renewcommand{\arraystretch}{1.15}
\begin{tabular}{ccccc}
\hline 
Datasets & Methods & Recall & Precision & F1-measure\tabularnewline

\hline
\multirow{9}{*}{Devign} & TextCNN & 62.47\%  & 59.76\% & 61.08\%
\tabularnewline
 & RoBERTa & 46.61\%  & 64.78\% & 54.22\%
 \tabularnewline
 & CodeBERT & 60.48\%  & 60.00\% & 60.24\%
 \tabularnewline
 & GraphCodeBERT & 55.30\%  & 59.16\% & 57.17\%
 \tabularnewline
 & Devign & 58.80\%  & 58.95\% & 58.88\%
 \tabularnewline
 & ReGVD & 63.35\%  & 56.95\% & 59.98\%
 \tabularnewline
 & CodeT5 & 66.37\%  & 61.07\% & 63.61\%
 \tabularnewline
\cline{2-5} \cline{3-5} \cline{4-5} \cline{5-5} 
 & \ourapp-R (ours) & 81.35\%  & 56.47\% & \textbf{66.67\% }\textcolor{blue}{($\uparrow$3.06\%)}
 \tabularnewline
 & \ourapp-T (ours) & 78.73\%  & 57.61\% & \underline{66.53\% }\textcolor{blue}{($\uparrow$2.92\%)}
 \tabularnewline

\hline 
\multirow{9}{*}{ReVeal} & TextCNN & 33.48\%  & 47.53\% & 39.29\%\tabularnewline
 & RoBERTa & 27.83\%  & 68.82\% & 39.63\%
 \tabularnewline
 & CodeBERT & 30.43\%  & 64.81\%  & 41.42\%
 \tabularnewline
 & GraphCodeBERT & 30.00\%  & 69.70\% & 41.95\%
 \tabularnewline
 & Devign & 24.78\%  & 54.29\% & 34.03\%
 \tabularnewline
 & ReGVD & 27.39\%  & 58.33\% & 37.28\%
 \tabularnewline
 & CodeT5 & 30.87\%  & 65.14\% & 41.89\% 
 \tabularnewline
\cline{2-5} \cline{3-5} \cline{4-5} \cline{5-5} 
 & \ourapp-R (ours) & 53.04\%  & 37.65\% & \underline{44.04\% }\textcolor{blue}{($\uparrow$2.09\%)}
 \tabularnewline
 & \ourapp-T (ours) & 49.13\%  & 48.92\% & \textbf{49.02\% }\textcolor{blue}{($\uparrow$7.07\%)}
 \tabularnewline

\hline 
\multirow{9}{*}{D2A} & TextCNN & 76.95\%  & 57.52\% & 65.83\%\tabularnewline
 & RoBERTa & 81.82\%  & 56.88\% & 67.11\%
 \tabularnewline
 & CodeBERT & 74.03\%  & 59.38\% & 65.90\%
 \tabularnewline
 & GraphCodeBERT & 83.77\%  & 55.13\% & 66.49\%
 \tabularnewline
 & Devign & 75.00\%  & 58.78\% & 65.91\%
 \tabularnewline
 & ReGVD & 77.92\%  & 60.15\% & 67.89\%
 \tabularnewline
 & CodeT5 & 75.32\%  & 59.34\% & 66.38\%
 \tabularnewline
\cline{2-5} \cline{3-5} \cline{4-5} \cline{5-5} 
 & \ourapp-R (ours) & 89.29\% & 57.77\% & \underline{70.15\% }\textcolor{blue}{($\uparrow$2.26\%)}
 \tabularnewline
 & \ourapp-T (ours) & 90.91\%  & 57.85\% & \textbf{70.71\% }\textcolor{blue}{($\uparrow$2.82\%)}
 \tabularnewline

\hline 
\end{tabular}}
\vspace{-1mm}
\end{table}

In general, our approach achieves significantly higher performances than the baselines, ranging from 3.06\% to 12.45\% with \ourapp-R and from \textit{2.92\% to 12.31\%} with \ourapp-T on the F1-measure on the Devign dataset. To the Reveal dataset, our approach also achieves much higher performances than the baselines, ranging from \textit{2.09\% to 10.01\%} with \ourapp-R and from \textit{7.07\% to 14.99\%} with \ourapp-T on the F1-measure. On the other hand, our approach achieves markedly higher performances than the baselines, ranging from 2.26\% to 4.32\% with \ourapp-R and from 2.82\% to 4.88\% with \ourapp-T on the F1-measure on the D2A dataset.

\begin{figure}[ht]
\vspace{0mm}
\begin{centering}
\includegraphics[width=0.86\columnwidth]{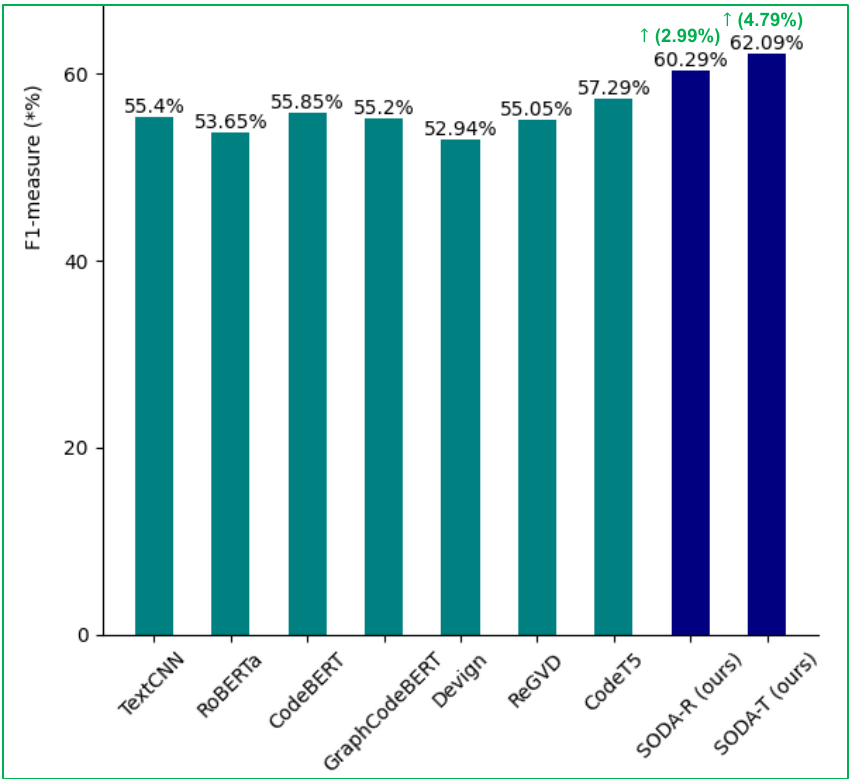}
\par\end{centering}\vspace{0.5mm}
\caption{The average results for the F1-measure of our \ourapp~approach and the baselines across all datasets mentioned in Table \ref{tab:mainresults}. Our \ourapp-R and \ourapp-T methods obtain significantly higher performances on the F1-measure than the second best baseline around 2.99\% and 4.79\%, respectively.}\label{fig:averageresultsf1}
\vspace{0mm}
\end{figure}%

\begin{figure}[ht]
\vspace{0mm}
\begin{centering}
\includegraphics[width=0.86\columnwidth]{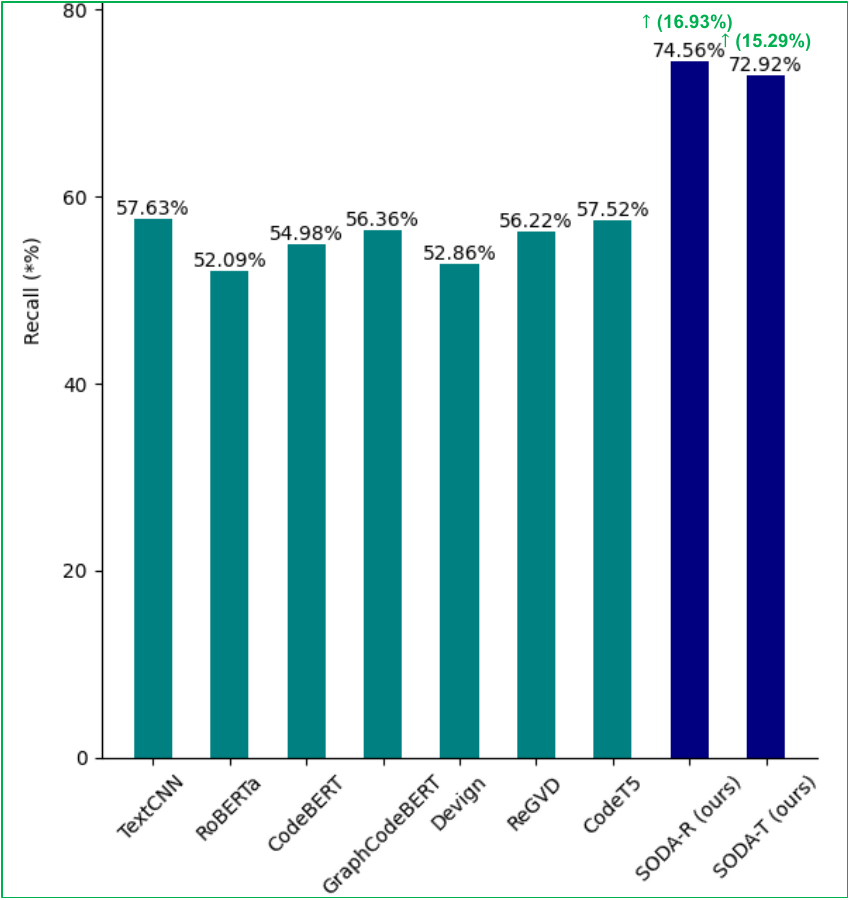}
\par\end{centering}\vspace{0.5mm}
\caption{The average results for the Recall measure of our \ourapp~approach and the baselines across all datasets mentioned in Table \ref{tab:mainresults}. Our \ourapp-R and \ourapp-T methods obtain much higher performances on the Recall measure than the second best baseline around 16.93\% and 15.29\%, respectively.}\label{fig:averageresultsrecall}
\vspace{0mm}
\end{figure}

In Figures \ref{fig:averageresultsf1} and \ref{fig:averageresultsrecall}, we present the average experimental results for Recall and F1-measure, respectively, of our \ourapp~approach and the baselines across three real-world, complex, and diverse datasets in Table \ref{tab:mainresults}. These average results again highlight the substantial advancement of our \ourapp~approach over the baselines. Specifically, for the F1-measure, our approach achieves significantly higher performance, ranging from \textit{2.99\% to 7.35\%} with \ourapp-R and from \textit{4.79\% to 9.15\%} with \ourapp-T compared to the baselines. Similarly, for the Recall measure, our method obtains significantly higher performance, ranging from \textit{16.93\% to 21.70\%} with \ourapp-R and from \textit{15.29\% to 20.06\%} with \ourapp-T over the baselines.

\vspace{5mm}
\colorbox{gray!20}{
\begin{minipage}{0.91\columnwidth}
\textbf{In conclusion for RQ1}: The experimental results in Table \ref{tab:mainresults} and Figures \ref{fig:averageresultsf1} and \ref{fig:averageresultsrecall} demonstrate the effectiveness and advancement of our \ourapp~approach. It achieves significantly higher performance on three real-world challenging source code datasets, particularly in terms of Recall and F1-measure. Specifically, on average, our \ourapp~approach outperforms the baselines from approximately 4.79\% (with \ourapp-T) and 16.93\% (with \ourapp-R)  for F1-measure and Recall, respectively.

\end{minipage}
}
\vspace{3mm}

\vspace{3mm}
\textbf{RQ2: \rqtwo}
\vspace{0mm}

\vspace{0mm}
\paragraph{\textbf{Approach}}

We investigate the performance of our \ourapp~approach in cases with and without using the enhancement of semantic and syntactic relationships from the source code data. This enhancement is achieved via distilled knowledge from the trained-teacher-A and trained-teacher-B models. Specifically, we examine the performance of our approach in four cases:

\vspace{1mm}
\begin{itemize}
    \item (i) Using only the Student-S model, denoted as \textit{\textbf{\ourapp-w/o-teacherAB}} or \textit{\textbf{\ourapp-w/oAB}} for short. \vspace{1mm}
    \item (ii) Using the Student-S model with distilled knowledge from the trained-teacher-A model (enhancing the semantic relationship), denoted as \textit{\textbf{\ourapp-w/o-teacherB}} or \textit{\textbf{\ourapp-w/oB}} for short. \vspace{1mm}
    \item (iii) Using the Student-S model with distilled knowledge from the trained-teacher-B model (enhancing the syntactic relationship), denoted as \textit{\textbf{\ourapp-w/o-teacherA}} or \textit{\textbf{\ourapp-w/oA}} for short. \vspace{1mm}
    \item (iv) Using the Student-S model with distilled knowledge from both the trained-teacher-A (enhancing the semantic relationship) and trained-teacher-B models (enhancing the syntactic relationship), denoted as \textit{\textbf{\ourapp-w-teacherAB}} or \textit{\textbf{\ourapp-wAB}} or \textit{\textbf{\ourapp~}}for short.
\end{itemize}

\vspace{1mm}
\paragraph{\textbf{Results}}

\vspace{3mm}
\begin{table}[h]
\vspace{0mm}
\caption{The results for Recall, Precision, and F1-measure of our approach in cases using and without using the enhancement of the semantic and syntactic relationships from the source code data via the distilled knowledge from the trained teacher-A and teacher-B models. We denote our approach as \ourapp-R and \ourapp-T when using RoBERTa and CodeT5, respectively, as the backbone for the Student-S model. The numbers highlighted in blue represent the improvements of cases (ii), (iii), and (iv) compared to case (i).}\label{tab:ablationstudies}
\vspace{1mm}
\centering{}
\resizebox{1.0\columnwidth}{!}{
\renewcommand{\arraystretch}{1.25}
\begin{tabular}{ccccc}
\hline 
Datasets & Methods & Recall & Precision & F1-measure\tabularnewline
\hline 
\multirow{4}{*}{Devign} & \ourapp-R-w/oAB & 46.61\% & 64.78\% & 54.22\%
\tabularnewline
 & \ourapp-R-w/oB & 81.35\% & 54.75\% & 65.45\% \textcolor{blue}{($\uparrow$11.23\%)}
 \tabularnewline
 & \ourapp-R-w/oA & 77.21\% & 57.07\% & 65.63\% \textcolor{blue}{($\uparrow$11.41\%)}
 \tabularnewline
\cline{2-5} \cline{3-5} \cline{4-5} \cline{5-5} 
 & \ourapp-R-wAB & 81.35\% & 56.47\% & \textbf{66.67\% }\textcolor{blue}{($\uparrow$\textbf{12.45\%})}
 \tabularnewline
\hline 

\multirow{4}{*}{ReVeal} & \ourapp-R-w/oAB & 27.83\% & 68.82\% & 39.63\%
\tabularnewline
 & \ourapp-R-w/oB & 46.09\% & 38.69\% & 42.06\% \textcolor{blue}{($\uparrow$2.43\%)}
 \tabularnewline
 & \ourapp-R-w/oA & 47.83\% & 37.67\% & 42.15\% \textcolor{blue}{($\uparrow$2.52\%)}
 \tabularnewline
\cline{2-5} \cline{3-5} \cline{4-5} \cline{5-5} 
 & \ourapp-R-wAB & 53.04\% & 37.65\% & \textbf{44.04\% }\textcolor{blue}{($\uparrow$\textbf{4.41\%})}
 \tabularnewline
\hline 

\multirow{4}{*}{D2A} & \ourapp-T-w/oAB & 81.82\% & 56.88\% & 67.11\%
\tabularnewline
 & \ourapp-R-w/oB & 76.30\% & 60.26\% & 67.34\% \textcolor{blue}{($\uparrow$0.23\%)}
 \tabularnewline
 & \ourapp-R-w/oA & 81.49\% & 59.34\% & 68.67\% \textcolor{blue}{($\uparrow$1.56\%)}
 \tabularnewline
\cline{2-5} \cline{3-5} \cline{4-5} \cline{5-5} 
 & \ourapp-R-wAB & 89.29\% & 57.77\% & \textbf{70.15\% }\textcolor{blue}{($\uparrow$\textbf{3.04\%})}
 \tabularnewline
\hline 
\hline

\multirow{4}{*}{Devign} & \ourapp-T-w/oAB & 66.37\% & 61.07\% & 63.61\%
\tabularnewline
 & \ourapp-T-w/oB & 78.09\% & 56.71\% & \textbf{65.71\%} \textcolor{blue}{($\uparrow$2.10\%)}
 \tabularnewline
 & \ourapp-T-w/oA & 73.55\% & 59.40\% & 65.72\% \textcolor{blue}{($\uparrow$2.11\%)}
 \tabularnewline
\cline{2-5} \cline{3-5} \cline{4-5} \cline{5-5} 
 & \ourapp-T-wAB & 78.73\% & 57.61\% & \textbf{66.53\% }\textcolor{blue}{($\uparrow$\textbf{2.92\%})}
 \tabularnewline
\hline 

\multirow{4}{*}{ReVeal} & \ourapp-T-w/oAB & 30.87\% & 65.14\% & 41.89\%
\tabularnewline
 & \ourapp-T-w/oB & 45.22\% & 52.26\% & 48.48\% \textcolor{blue}{($\uparrow$6.59\%)}
 \tabularnewline
 & \ourapp-T-w/oA & 44.78\% & 53.09\% & 48.58\% \textcolor{blue}{($\uparrow$6.69\%)}
 \tabularnewline
\cline{2-5} \cline{3-5} \cline{4-5} \cline{5-5} 
 & \ourapp-T-wAB & 49.13\% & 48.92\% & \textbf{49.02\% }\textcolor{blue}{($\uparrow$\textbf{7.13\%})}
 \tabularnewline
\hline 

\multirow{4}{*}{D2A} & \ourapp-T-w/oAB & 75.32\% & 59.34\% & 66.38\%
\tabularnewline
 & \ourapp-T-w/oB & 80.52\% & 57.54\% & 67.12\% \textcolor{blue}{($\uparrow$0.74\%)}
 \tabularnewline
 & \ourapp-T-w/oA & 90.58\% & 56.59\% & 69.66\% \textcolor{blue}{($\uparrow$3.28\%)}
 \tabularnewline
\cline{2-5} \cline{3-5} \cline{4-5} \cline{5-5} 
 & \ourapp-T-wAB & 90.91\% & 57.85\% & \textbf{70.71\% }\textcolor{blue}{($\uparrow$\textbf{4.33\%})}
 \tabularnewline
\hline 
\end{tabular}}
\vspace{0mm}
\end{table}

The experimental results in Table \ref{tab:ablationstudies} demonstrate the effectiveness of enhancing semantic and syntactic relationships in source code data via the distilled knowledge from the trained-teacher-A and trained-teacher-B models. The model performance, especially the F1-measure, also significantly increases with either semantic or syntactic enhancement compared to the case without these enhancements. 

\vspace{1mm}
The best F1-measure and Recall results are achieved when both semantic and syntactic relationships are utilized. In particular, in this case, for F1-measure, our \ourapp-R-wAB method achieves a significantly higher performance of 12.45\%, 4.41\%, and 3.04\%  for the Devign, Reveal, and D2A datasets, respectively. On the other hand, our \ourapp-T-wAB method results in performance increases of 2.92\%, 7.13\%, and 4.33\% for the Devign, Reveal, and D2A datasets, correspondingly.

\vspace{2mm}
\colorbox{gray!20}{
\begin{minipage}{0.91\columnwidth}
\textbf{In conclusion for RQ2:} The experimental results in Table \ref{tab:ablationstudies} demonstrate the effectiveness of utilizing and enhancing semantic and syntactic relationships in improving the capability of our \ourapp~approach for software vulnerability detection (SVD), particularly in terms of F1-measure and Recall, the two key and prioritized metrics in the field.
\end{minipage}
}
\vspace{0mm}

\vspace{1mm}
\textbf{RQ3: \rqthree}
\vspace{0mm}

\paragraph{\textbf{Approach}}
We investigate the effect of utilizing different hierarchical structures from the source code data on the performance of our \ourapp~approach including (i) ASTs (furnishing comprehensive syntax details of the source code) and (ii) DFGs (representing the \quotes{where-the-value-comes-from} relationships between variables, allowing us to trace the flow of data within the code), utilized in the teacher-B model.

\vspace{1mm}
\paragraph{\textbf{Results}}

As shown in Table \ref{tab:astdfg}, while the performance of our \ourapp\ approach varies across the used datasets when using ASTs and DFGs, it consistently achieves state-of-the-art results in F1-measure and Recall compared to the baselines (as detailed in Table \ref{tab:mainresults}).

Along with the results depicted in Tables \ref{tab:mainresults} and \ref{tab:ablationstudies}, the experimental results shown in Table \ref{tab:astdfg}, further demonstrate the effectiveness of leveraging hierarchical code structures (i.e., ASTs or DFGs) in enhancing the performance of our \ourapp~approach for solving the SVD problem. Incorporating the hierarchical structure relationships through the distilled knowledge from the trained-teacher-B model, along with the semantic relationships through the distilled knowledge from the trained-teacher-A model, from the source code data helps our \ourapp~approach consistently achieves substantially higher results in F1-measure and Recall compared to the baselines.

\begin{table}[h]
\caption{The experimental results for Recall, Precision, and F1-measure of our approach when utilizing AST and DFG structures from the source code data. We denote our approach as \ourapp-R-AST and \ourapp-R-DFG when using RoBERTa as the backbone for the Student-S model with ASTs and DFGs for the Teacher-B model, respectively. Similarly, \ourapp-T-AST and \ourapp-T-DFG refer to our approach when using CodeT5 as the backbone for the Student-S model with ASTs and DFGs for the Teacher-B model, respectively.}\label{tab:astdfg}
\vspace{-1mm}
\centering{}
\resizebox{0.95\columnwidth}{!}{
\renewcommand{\arraystretch}{1.25}
\begin{tabular}{ccccc}
\hline 
Datasets & Methods & Recall & Precision & F1-measure\tabularnewline
\hline 


\multirow{2}{*}{Devign} & \ourapp-R-AST & 81.35\%  & 56.47\% & \underline{66.67\%}\tabularnewline
 & \ourapp-R-DFG & 79.36\%  & 56.88\% & 66.27\%\tabularnewline
& \ourapp-T-AST & 78.73\%  & 57.61\% & 66.53\%\tabularnewline
 & \ourapp-T-DFG & 80.16\%  & 57.42\% & \textbf{66.91\%}\tabularnewline
\hline 

\multirow{2}{*}{ReVeal} & \ourapp-R-AST & 53.04\%  & 37.65\% & 44.04\%\tabularnewline
 & \ourapp-R-DFG & 44.35\%  & 40.64\% & 42.41\%\tabularnewline
 & \ourapp-T-AST & 49.13\%  & 48.92\%  & \underline{49.02\%}\tabularnewline
 & \ourapp-T-DFG & 52.40\%  & 46.33\% & \textbf{49.18\%}\tabularnewline
\hline 

\multirow{2}{*}{D2A} & \ourapp-R-AST & 89.29\%  & 57.77\% & \underline{70.15\%}\tabularnewline
 & \ourapp-R-DFG & 85.39\%  & 57.93\% & 69.03\%\tabularnewline
 & \ourapp-T-AST & 90.91\%  & 57.85\% & \textbf{70.71\%}\tabularnewline
 & \ourapp-T-DFG & 91.56\% & 56.06\% & 69.54\%\tabularnewline
\hline
\end{tabular}}
\vspace{-2mm}
\end{table}

\begin{figure*}[ht]
\vspace{-2mm}
\begin{centering}
\includegraphics[width=1.0\textwidth]{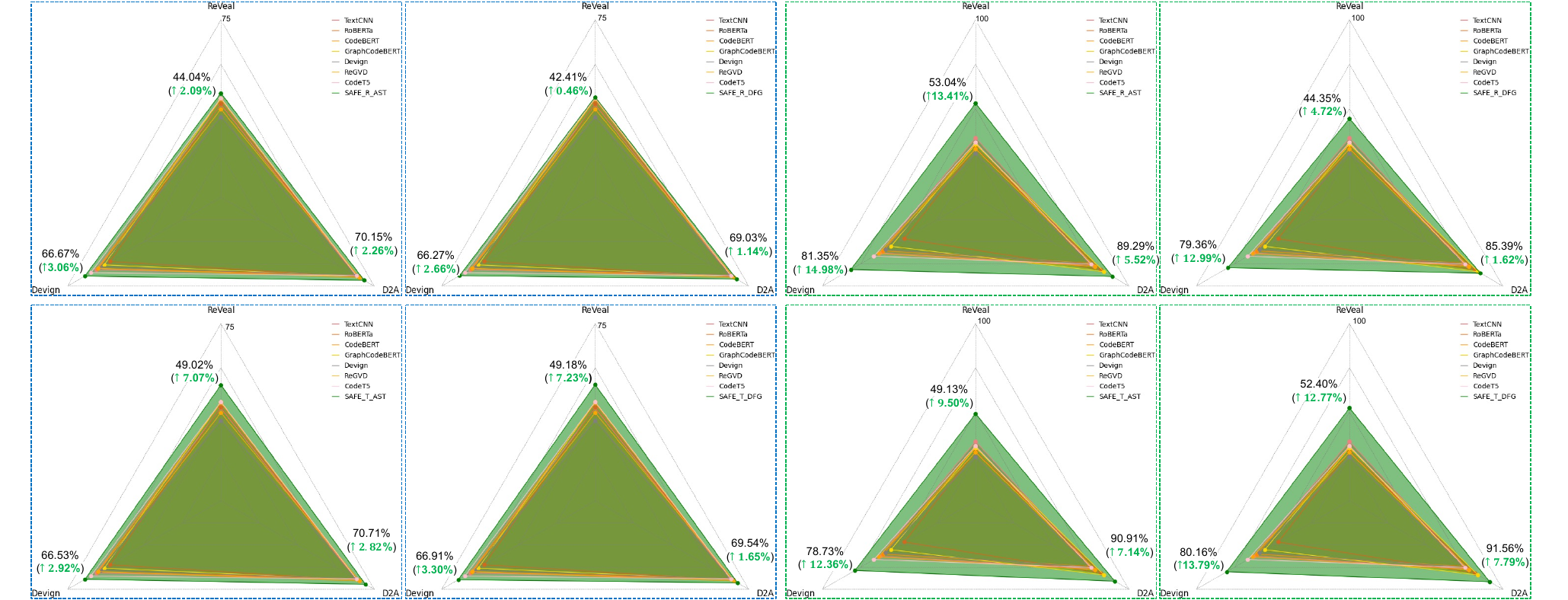}
\par\end{centering}\vspace{1mm}
\caption{A comparison for the F1-measure (four first sub-figures with dotted blue borders) and Recall (four second sub-figures with dotted green borders) of our \ourapp~approach's variants (i.e., \ourapp-R-AST, \ourapp-R-DFG, \ourapp-T-AST, and \ourapp-T-DFG) when alternatively utilizing ASTs and DFGs from the source code data for the Teacher-B model as well as leveraging RoBERTa and CodeT5 for the Student-S model (shown in Table \ref{tab:astdfg}) and the baselines (mentioned in Table \ref{tab:mainresults}) across the used datasets. For each variant (method) of our \ourapp~approach on each dataset, we display its corresponding performance, with the improvements over the second-best baseline shown as green numbers in brackets.}\label{fig:rq3resultsf1recall}
\vspace{-2mm}
\end{figure*}

In Figure \ref{fig:rq3resultsf1recall}, we further compare the performance of our \ourapp~approach, alternatively utilizing the hierarchical code structures (i.e., ASTs and DFGs) from the source code data for the Teacher-B model and leveraging RoBERTa and CodeT5 as the backbone for the Student-S model, against the baselines. These comparisons are made using Radar charts across three used datasets (Reveal, D2A, and Devign) with the performance measured by F1-measure and Recall.

Note that in each sub-figure, a specific variant (method) of our \ourapp~approach is represented in green. It is evident that in each sub-figure, \textit{the green area dominates the other color domains (representing the baselines) and covers a noticeably larger region}. This is especially pronounced in Recall for all variants and in the F1-measure for the variants utilizing CodeT5 as the backbone for the Student-S model.

\vspace{3mm}
\colorbox{gray!20}{
\begin{minipage}{0.91\columnwidth}
\textbf{In conclusion for RQ3:} The results in Table \ref{tab:astdfg} demonstrate the effectiveness of utilizing the hierarchical structures, i.e., ASTs and DFGs, along with the semantic relationships, in improving the capability of our \ourapp~approach for addressing the SVD problem. As a result, our \ourapp~approach achieves the best performance in the two key and prioritized metrics (i.e., F1-measure and Recall), significantly outperforming the baselines.
\end{minipage}
}
\vspace{1mm}

\vspace{0mm}
\subsection{\textbf{Threats to Validity}}\label{sec:threats}

\vspace{-0.5mm}
\paragraph{\textbf{Construct Validity}}
Key construct validity threats are if our assessments of the methods demonstrate the ability for software vulnerability detection (SVD). The main purpose of our \ourapp~approach is for solving the SVD problem in real-world, challenging, complex, and diverse source code datasets. To evaluate the performance of our \ourapp~approach and baselines, we use three main measures including Recall, Precision, and F1-measure (e.g., \cite{VulDeePecker2018,Li2018SySeVR,vannguyen2019dan,DevignZhou2019, d2a2021}). In the field of software vulnerability detection, the F1-measure can be considered the most important metric, with Recall having a higher priority than Precision \cite{ami2023false}.

\vspace{1.0mm}
\paragraph{\textbf{Internal Validity}}
Key internal validity threats are relevant to the choice of hyper-parameter settings (such as optimizer, learning rate, and the number of layers in deep neural networks). It is worth noting that finding a set of optimal hyper-parameter settings of deep neural networks is expensive due to a large number of trainable parameters. To train our \ourapp~approach, we only use the common and default values of hyper-parameters such as using Adam optimizer and the learning rate in $\{2e^{-5},5e^{-5}\}$. We also report the hyper-parameter settings in the released reproducible source code to support future replication studies.

\vspace{0.5mm}
\paragraph{\textbf{External Validity}}
Key external validity threats include whether our \ourapp~approach can generalize well to real-world, complex, and diverse source code vulnerabilities. We mitigated this problem by conducting our rigorous and extensive experiments on three real-world challenging source code datasets (i.e., ReVeal, D2A, and Devign) covering various real-world vulnerability types. The experimental results show the advancements of our \ourapp~approach over the baselines by a wide margin, particularly in the two key and prioritized metrics, F1-measure and Recall.

\vspace{3mm}
\vspace{-1mm}
\section{Conclusion}\label{sec:conclusion}

In this paper, we have successfully proposed an innovative deep learning-based approach for software vulnerability detection (SVD). In particular, our \ourapp~approach enhances the capabilities of large language models (LLMs) to learn and leverage the semantic and syntactic relationships (ie.., constituting crucial information in the formula of the associated source code data) from the source code data by elegantly utilizing the distilled knowledge from the teacher models via the distillation mechanism. This fusion allows our \ourapp~approach not only can acquire the fundamental knowledge from source code data (i.e., as a consequence of being built from a pre-trained large language model) but also can adeptly utilize the source code data crucial semantic and syntactic association, for effectively tackling the software vulnerability detection (SVD) problem. The rigorous and extensive experimental results on three real-world, complex, and diverse source code datasets demonstrate the effectiveness and advancement of our \ourapp~method compared to the seven effective and state-of-the-art baselines by a wide margin on F1-measure and Recall, the key and prioritized metrics in SVD.

\vspace{3mm}
\bibliographystyle{IEEEtran}
\bibliography{reference}

\vspace{3mm}
\section{Appendix}
\vspace{0mm}
\subsection{\textbf{Model's configurations}}

For the baselines of our \ourapp~approach, i.e., TextCNN \cite{textCNNKim2014}, RoBERTa \cite{LiuRoBERTa2019}, Devign \cite{DevignZhou2019}, CodeBERT \cite{CodeBERT2020}, GraphCodeBERT \cite{Guographcodebert2021}, CodeT5 \cite{wang-etal-2021-codet5}, and ReGVD \cite{ReGVD2022}, we use the architectures and all hyper-parameters suggested in the corresponding papers when applying for software vulnerability detection (SVD). In our proposed \ourapp~approach, we use a convolutional neural network for the teacher-A model, as used in TextCNN \cite{textCNNKim2014}, and a graph neural network for the teacher-B model, as used in ReGVD \cite{ReGVD2022}, due to their lightweight nature and effectiveness in learning semantic and syntactic relationships in source code data, respectively. For the student-S model, we utilize either of two effective and popular pre-trained large language models, i.e., RoBERTa or CodeT5 (i.e., for CodeT5, we only use its encoder part specializing for data representation learning). 

In our \ourapp~approach, we vary the trade-off hyper-parameters $\gamma$, $\delta$, and $\eta$ to set the weights for utilizing the pre-trained large language model backbone in the Student-S model and enhancing the semantic and syntactic relationships transferred from the trained-teacher-A and trained-teacher-B models. The hyper-parameter $\gamma$ is varied within the range of $\{0.3, 0.5, 0.7\}$, while $\delta$ is defined as $(1-\gamma) \times \kappa$ where $\kappa \in \{0.3, 0.5, 0.7\}$. Additionally, $\eta$ is set to $1-(\gamma+\delta$). This setup allows us to explore different configurations and their impact on the overall performance of our \ourapp~approach. Furthermore, the temperature $T$ is simply set to 1.

We run our experiments on a 13th Gen Intel(R) Core(TM) i9-13900KF having 24 CPU Cores at 3.00 GHz with 32GB RAM, integrated Gigabyte RTX 4090 Gaming OC 24GB.

\subsection{\textbf{Additional experiments}}

In this section, we further visualize the comparison between our \ourapp~approach (via its variants) and the baselines, focusing on the model performance in terms of the F1-measure and Recall metrics across three real-world challenging datasets used (i.e., Reveal, D2A, and Devign).

The variants of our \ourapp~approach include \ourapp-R-AST, \ourapp-R-DFG, \ourapp-T-AST, and \ourapp-T-DFG. These correspond to the use of hierarchical code structures (ASTs and DFGs) from the source code data for the Teacher-B model, along with two effective and popular pre-trained large language models (RoBERTa and CodeT5) as the backbone for the Student-S model. In short, we denote our approach as \ourapp-R-AST and \ourapp-R-DFG when using RoBERTa as the backbone for the Student-S model with ASTs and DFGs for the Teacher-B model, respectively. Similarly, \ourapp-T-AST and \ourapp-T-DFG refer to our approach when using CodeT5 as the backbone for the Student-S model with ASTs and DFGs for the Teacher-B model, respectively.

\begin{figure}[ht]
\vspace{1mm}
\begin{centering}
\includegraphics[width=0.9\columnwidth]{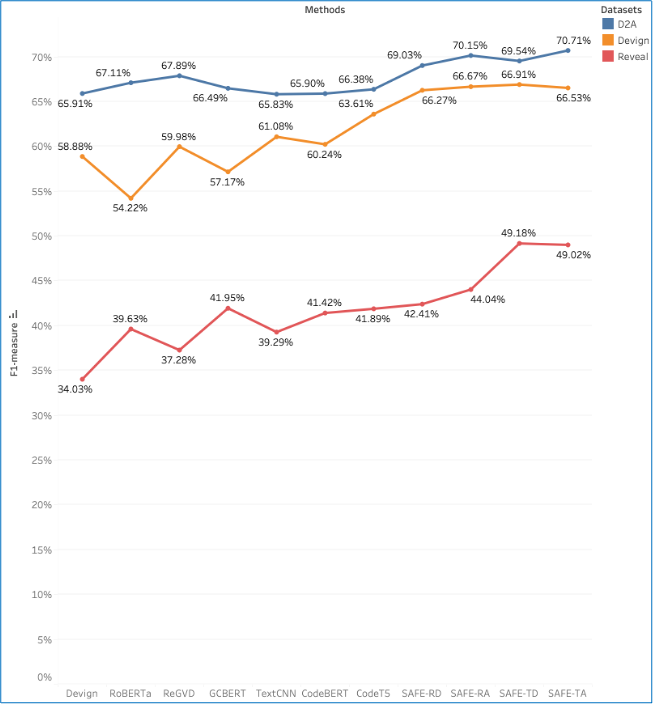}
\par\end{centering}\vspace{-1mm}
\caption{The experimental results for F1-measure of our \ourapp~approach's variants (i.e., \ourapp-R-AST, \ourapp-R-DFG, \ourapp-T-AST, and \ourapp-T-DFG) and the baselines across all datasets mentioned in Tables 1 and 3 (in the main paper), sorted ascending by sum of F1-measure within methods across three used datasets. We denote GraphCodeBERT, \ourapp-R-AST, \ourapp-R-DFG, \ourapp-T-AST, and \ourapp-T-DFG as GCBERT, \ourapp-RA, \ourapp-RD, \ourapp-TA, and \ourapp-TD, respectively, for short.}\label{fig:f1_tables1and3}
\vspace{1mm}
\end{figure}%

\begin{figure}[ht]
\vspace{1mm}
\begin{centering}
\includegraphics[width=0.9\columnwidth]{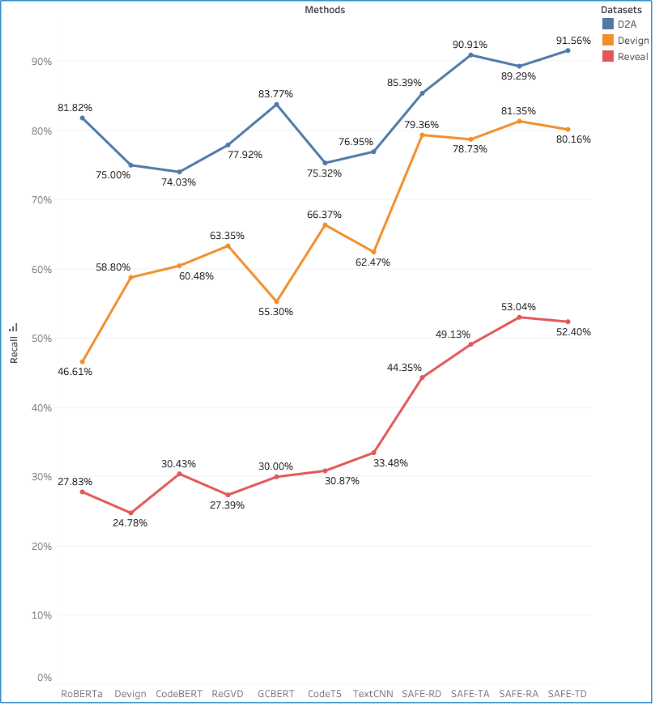}
\par\end{centering}\vspace{1mm}
\caption{The experimental results for Recall of our \ourapp~approach's variants (i.e., \ourapp-R-AST, \ourapp-R-DFG, \ourapp-T-AST, and \ourapp-T-DFG) and the baselines across all datasets mentioned in Tables 1 and 3 (in the main paper), sorted ascending by sum of Recall within methods across three used datasets. We denote GraphCodeBERT, \ourapp-R-AST, \ourapp-R-DFG, \ourapp-T-AST, and \ourapp-T-DFG as GCBERT, \ourapp-RA, \ourapp-RD, \ourapp-TA, and \ourapp-TD, respectively, for short.}\label{fig:recall_tables1and3}
\vspace{1mm}
\end{figure}%

As shown in Figures \ref{fig:f1_tables1and3} and \ref{fig:recall_tables1and3}, all variants of our \ourapp~approach (i.e., \ourapp-R-AST, \ourapp-R-DFG, \ourapp-T-AST, and \ourapp-T-DFG) achieve significantly higher F1-measure and Recall performances compared to the baselines across all datasets. Moreover, the model performance of our \ourapp~approach and baselines for F1-measure and Recall on the D2A dataset is better than on the Devign dataset, followed by the Reveal dataset.

\vspace{1mm}
Among the variants of our \ourapp~approach, \ourapp-T-AST and \ourapp-T-DFG achieve the highest performance in F1-measure and Recall on the D2A dataset, respectively, compared to other variants and the baselines. In contrast, \ourapp-T-DFG and \ourapp-R-AST perform best in F1-measure and Recall, respectively, on the Devign and Reveal datasets.

\vspace{1mm}
\subsection{\textbf{Qualitative evaluations}}
In the testing set of the Reveal dataset, there are 2,274 source code functions, of which 230 (10.1\%) are vulnerable. In the testing set of D2A dataset, there are 596 source code functions, with 303 (50.8\%) being vulnerable while in the testing set of the Devign dataset contains 2,732 source code functions, with 1,255 (45.9\%) of them vulnerable.

In this section, we present \textbf{some examples of vulnerable source code functions that our proposed \ourapp~approach successfully identifies, while all baseline methods fail to do so}. These examples are drawn from the testing set of three datasets used including Reveal, D2A, and Devign.

\begin{figure}[ht]
\vspace{2mm}
\begin{centering}
\includegraphics[width=0.99\columnwidth]{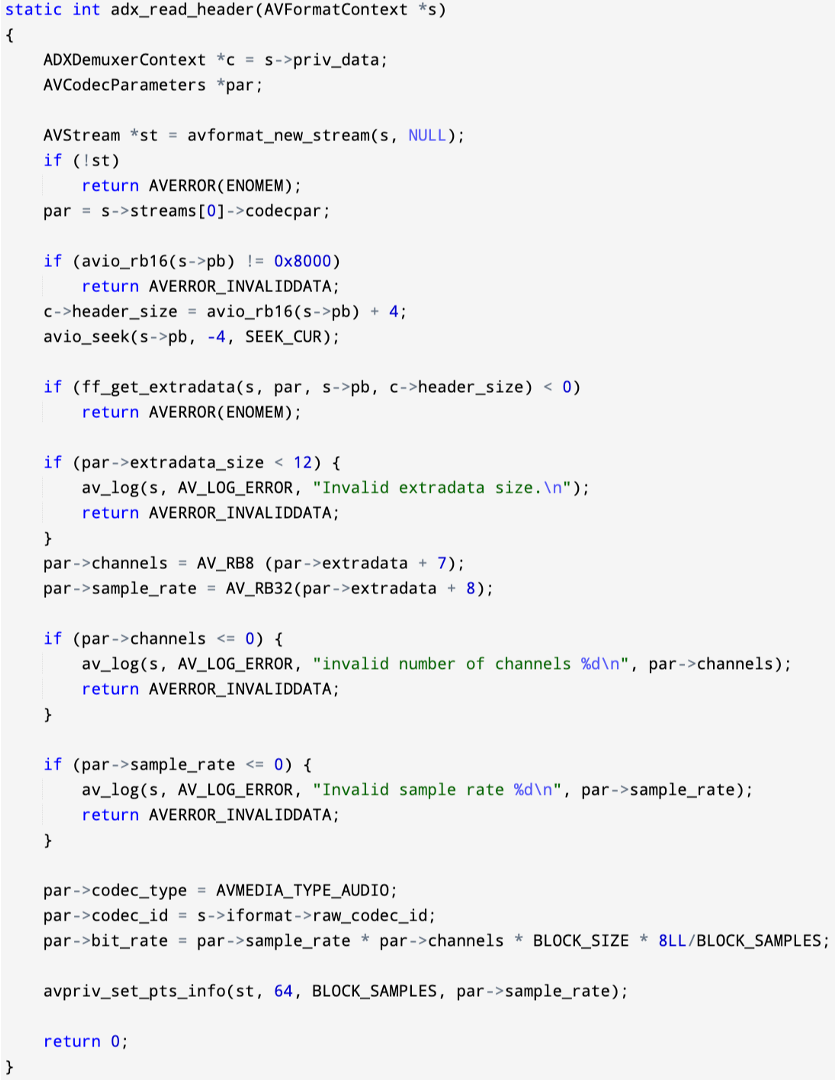}
\par\end{centering}\vspace{1mm}
\caption{A vulnerable source code sample from the D2A dataset that our proposed \ourapp~approach successfully identifies, while all baseline methods fail to do so.}\label{fig:eval_d2a}
\vspace{0mm}
\end{figure}%

The function \textit{adx\_read\_header}, shown in Figure \ref{fig:eval_d2a}, is from the D2A dataset and is part of a demuxer for handling ADX (ADPCM Dialogic eXtended) audio files in a multimedia framework, such as Libav, FFmpeg, or a similar library. However, there are some potential vulnerabilities in the function, for example:

\begin{itemize}

\item \textbf{Potential Buffer Overflow}: The function reads \textit{c$\rightarrow$header\_size} bytes of extradata using \textit{ff\_get\_extradata} but does not ensure that this size is within the buffer's limits. If \textit{c$\rightarrow$header\_size} is larger than the available data in \textit{s$\rightarrow$pb}, it can lead to out-of-bounds reads. Buffer overflow can occur if \textit{c$\rightarrow$header\_size} exceeds the actual buffer size, leading to potential crashes or data corruption.

\item \textbf{Resource Leaks}: The function allocates memory for \textit{st} and \textit{par} but does not release the memory in case of errors, leading to memory leaks.

\end{itemize}

\begin{figure}[ht]
\vspace{2mm}
\begin{centering}
\includegraphics[width=0.9\columnwidth]{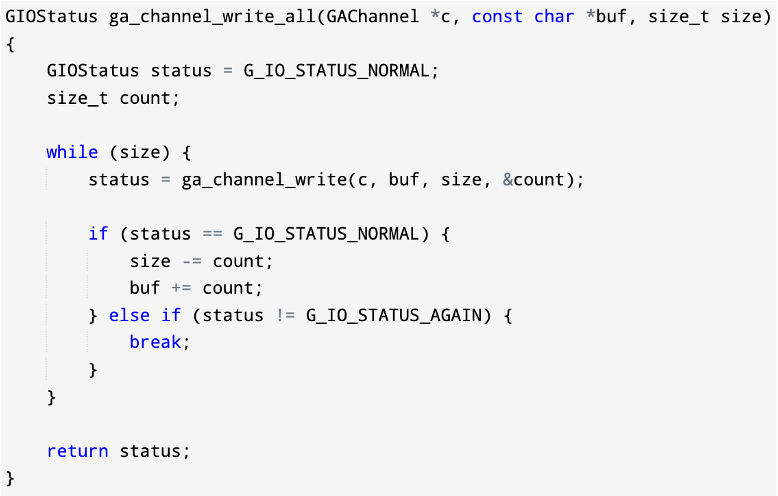}
\par\end{centering}\vspace{1mm}
\caption{A vulnerable source code sample from the Devign dataset that our proposed \ourapp~approach successfully identifies, while all baseline methods fail to do so.}\label{fig:eval_devign}
\vspace{0mm}
\end{figure}%

\vspace{3mm}
The function \textit{ga\_channel\_write\_all}, shown in Figure \ref{fig:eval_devign}, is from the Devign dataset and is designed to write a specified amount of data to a communication channel, represented by GAChannel, in a manner that ensures all the data is written. It uses the \textit{ga\_channel\_write} function to handle the actual write operations. However, there are some potential vulnerabilities in the function, for example:

\begin{itemize}

\item \textbf{Buffer Overrun}: The function does not check if \textit{count} is greater than size, which could lead to an overrun if \textit{ga\_channel\_write} returns an unexpectedly large value for \textit{count}. Buffer overruns can lead to memory corruption, crashes, or security vulnerabilities such as buffer overflows.

\item \textbf{Unbounded Loop}: The function contains a \textit{while (size)} loop that continues as long as size is non-zero. If \textit{ga\_channel\_write} always returns \textit{G\_IO\_STATUS\_AGAIN}, the loop will continue indefinitely. This can lead to a denial-of-service (DoS) condition, where the application becomes unresponsive or consumes excessive CPU resources due to the infinite loop.

\item \textbf{Lack of Error Handling}: The function breaks the loop if status is not \textit{G\_IO\_STATUS\_AGAIN}, but it does not provide specific handling or logging for different error statuses. Inadequate error handling can lead to unexpected behavior, making it difficult to diagnose and recover from errors.

\end{itemize}

\begin{figure}[ht]
\vspace{0mm}
\begin{centering}
\includegraphics[width=0.98\columnwidth]{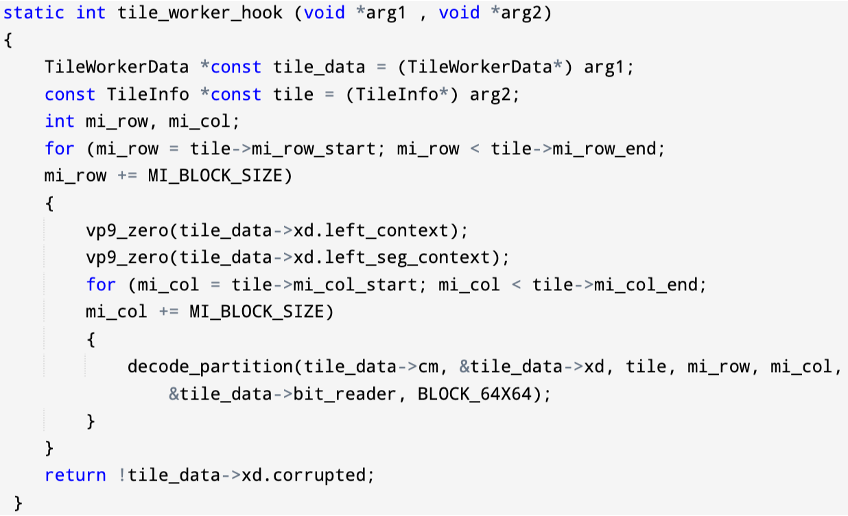}
\par\end{centering}\vspace{1mm}
\caption{A vulnerable source code sample from the Reveal dataset that our proposed \ourapp~approach successfully identifies, while all baseline methods fail to do so.}\label{fig:eval_reveal}
\vspace{0mm}
\end{figure}

\vspace{3mm}
The function \textit{tile\_worker\_hook}, shown in Figure \ref{fig:eval_reveal}, is from the Reveal dataset. This function is part of a video decoding process, likely related to decoding video tiles in a VP9 video codec. However, there are some potential vulnerabilities in the function, for example:

\begin{itemize}

\item \textbf{Buffer Overflows}: The loops iterate based on values from the tile structure (\textit{tile$\rightarrow$mi\_row\_start}, \textit{tile$\rightarrow$mi\_row\_end}, \textit{tile$\rightarrow$mi\_col\_start}, \textit{tile$\rightarrow$mi\_col\_end)}, which are not checked for correctness. If these values are manipulated (e.g., through a maliciously crafted tile), the loops could exceed the bounds of the memory allocated for \textit{tile\_data}, leading to buffer overflows.

\item \textbf{Unchecked Function Calls}: The function calls \textit{vp9\_zero} and \textit{decode\_partition} without checking their return values or ensuring that they execute successfully. If these functions fail or behave unexpectedly, the state of \textit{tile\_data} and other structures might be compromised.

\end{itemize}

Note that in these vulnerable functions mentioned in Figures (\ref{fig:eval_d2a}, \ref{fig:eval_devign}, and \ref{fig:eval_reveal}), we assume that the input arguments have already been validated. If not, these functions also face the Null Pointer Dereference vulnerability. Specifically, these functions do not validate their input arguments. If these pointers are NULL or point to invalid data, it can lead to undefined behavior or crashes. Without proper validation, the functions can potentially access invalid memory, leading to buffer overflows or other memory corruption issues.

\end{document}